\documentclass[12pt]{iopart}
\usepackage{graphicx}
\usepackage{iopams}

\usepackage{bm}

\begin{document}
\title [PrCoAsO and NdCoPO]{Local electromagnetic properties of magnetic pnictides: A comparative study probed by NMR measurement}
\author{M. Majumder\dag\footnote[1]{To whom correspondence should be addressed (mayukh.cu@gmail.com)}, K. Ghoshray\dag, A. Ghoshray\dag, A. Pal\ddag, V. P. S. Awana\ddag}

\address{\dag\ ECMP Division, Saha Institute of Nuclear Physics, 1/AF Bidhannagar,
Kolkata-700064, India}
\address{\ddag\ Quantum Phenomenon and Applications Division, National Physical Laboratory (CSIR), New Delhi-110012, India}

\date{\today}
\begin{abstract}

$^{75}$As and $^{31}$P NMR studies are performed in PrCoAsO and NdCoPO respectively. The Knight shift data in PrCoAsO indicate the presence of an  antiferromagnetic interaction between the 4$f$ moments along the $c$ axis in the ferromagnetic state of Co 3$d$ moments. We propose a possible spin structure in this system. The $^{75}$As quadrupolar coupling constant, $\nu_Q$ increases continuously with the decrease of temperature and is found to vary linearly with the intrinsic spin susceptibility, $K_{iso}$. This indicates a possibility of the presence of a coupling between charge density and spin density fluctuations. Further, $^{31}$P NMR Knight shift and spin lattice relaxation rate ($1/T_1$) in the paramagnetic state of NdCoPO indicate that the differences between LaCoPO and NdCoPO with SmCoPO are due to the decrement of inter layer separation and not due to the moments of 4$f$ electrons. Nuclear spin lattice relaxation time ($T_1$) in NdCoPO shows weak anisotropy at 300 K. Using self consistent renormalization (SCR) theory of itinerant ferromagnet, it is shown that in the $ab$ plane, the spin fluctuations are three dimensional ferromagnetic in nature. From SCR theory the important spin fluctuation parameters ($T_0$, $T_A$, $\bar{F}_1$) are evaluated. The similarities and dissimilarities of the NMR results in As and P based systems, with different rare earths have also been discussed.

\end{abstract}
 \ead{mayukh.cu@gmail.com}
\maketitle

\section{Introduction}
The main attraction of the layered FeAs based superconductors to the condensed matter physicists is due to their unconventional nature and the presence of several competing interactions. These superconductors are grouped in several families \cite{Johnston10}. In 1111 and 122 families, superconductivity can be achieved by Co doping in place of iron \cite{sefat08,2sefat08}. Whereas, in LaCo$_2$B$_2$, which also belongs to 122 family, superconductivity appears upon Y doping at the La site \cite{Mizoguchi11}. Thus it is emerged that Co is playing different role in different systems of the same type of crystal structure. Therefore, the non superconducting Co based members in 1111 and 122 family should be explored systematically with microscopic tools. This may help to understand the physics behind the emergence of superconductivity in non-doped cases. Apart from this fact, the novel and complex magnetic properties of the Co based magnetic systems \emph{RE}CoAsO/\emph{RE}CoPO (where \emph{RE} is rare-earth ion) belonging to 1111 family, are also of interest because of the 4$f$-3$d$ interplay \cite{Rajib10,Awana11}. Surprisingly, in \emph{RE}FeAsO series no such dominant effect of the interplay of 3$d$-4$f$ interaction was observed in magnetic properties.

Recently it was shown from magnetization and specific heat data that Sm/NdCoPO undergo three magnetic transitions i.e. $T_{C,Co}$ (80 K), the Sm$^{4f}$-Co$^{3d}$ and Nd$^{4f}$-Co$^{3d}$ inter-played antiferromagnetic (AFM) transition below 20 K and finally Sm$^{3+}$ and Nd$^{3+}$ spins individual AFM transitions at 5.4 K and below 2 K respectively \cite{Awana11}. Again in \emph{RE}CoAsO series La, Ce, and Pr show PM to FM transition, Sm, Nd, and Gd show PM $\rightarrow$ FM $\rightarrow$ AFM transition \cite{Awana11,Ohta09}. Furthermore, in SmCoAsO and NdCoAsO a second AFM transition only due to rare earth ion as in NdCoPO and SmCoPO was also reported \cite{Awana10,McGuire10}. The main difference between P and As based compounds is that the FM transition temperature ($T_c$) increases from La to Ce and for Pr, Nd, Sm it is almost same in \emph{RE}CoAsO series whereas in \emph{RE}CoPO family $T_c$ increases progressively as we go down the series from La to Sm. In each case, the lattice volume decreases across the series. In general, with the application of chemical or physical pressure, $T_c$ decreases due to the increment of density of state (DOS) at the Fermi level (magneto-volume effect). However, due to the lattice size decrement, the three dimensionality of the magnetic interaction may enhance causing an increment of $T_c$ and also T$_N$. These point towards the active role of the competing phenomena governing the actual ground state and most importantly the role of competing 4$f$ and 3$d$ moments makes these families a novel and complex one in the field of magnetism.

Our earlier $^{31}$P and $^{139}$La NMR measurements in LaCoPO (quasi 2D fermi surface) revealed that the spin fluctuation of 3$d$ electrons in the paramagnetic (PM) state is basically two dimensional (2D) in nature with non negligible three dimensional (3D) part and it is 3D in the ferromagnetic (FM) state \cite{Majumder09, Majumder10}. In SmCoPO, we have observed that the 3$d$-spin fluctuations in the $ab$ plane is primarily of 2D FM in nature, while along the $c$-axis, a signature of a weak 2D AFM spin fluctuations superimposed on weak FM spin-fluctuations, even in a field of 7 T and far above $T_N$, was observed. The interaction between Sm-4$f$ and Co-3$d$ spin fluctuations, has a contribution for the development of weak AFM spin fluctuations along $c$-axis at a temperature far above the AFM transition temperature of the Co-3$d$ spins below their FM transition \cite{Majumder12}. The moment of Sm 4$f$ and the separation between the $ab$ planes are less in case of SmCoPO than in NdCoPO. So, it would be interesting to study NdCoPO to understand the role of the change of the inter-plane separation and the moment of 4$f$ electrons on its complex magnetism. Further the nature and the dimensionality of 3$d$ spin-fluctuations within the $ab$ plane in NdCoPO may also be compared with that in SmCoPO and LaCoPO.

Recent muon-spin rotation and relaxation studies in PrCoAsO indicated a further change in the magnetic state below the Co FM transition ($T_C$$\sim$ 75 K) \cite{Sugiyama11}. In contrast, in PrCoPO, no such change was reported \cite{Prando12}. Since NMR is a sensitive tool to probe the magnetic interactions, we intend to study the magnetic behavior of PrCoAsO and NdCoPO using $^{75}$As and $^{31}$P NMR respectively. Moreover, for  NdCoPO, till now, no results on microscopic measurements are reported. Also the $^{75}$As NMR results are reported for the first time in this paper. Finally, the present results are compared with those in La based analogue, in order to understand the role of 4$f$ electrons in As and P based series.

\section{Experimental}
For the synthesis of polycrystalline sample of PrCoAsO, at first, ingot of Pr metal (Alfa Aesar, 99.9 \%) is cut into small pieces and grounded to make tiny pieces. It is then mixed with As powder(Alfa Aesar, 99.99 \%) and the mixture is kept inside a silica tube to prepare PrAs. The whole process is done in an argon atmosphere. The silica tube is evacuated and sealed and then carefully fired in a furnace at 550 $^\circ$C for 5 h and then at 800 $^\circ$C for 12 h. The obtained powder of PrAs is mixed with the powders of CoO (Alfa Aesar, 99.999 \%) in stoichiometric ratio and ground well in argon to avoid oxidation. The pelletized mixture of PrAs and CoO is then kept into an evacuated silica tube and fired at 1100 $^\circ$°C for 12 h. The powder sample is characterized by using x-ray diffraction technique with CuK$\alpha$ radiation at room temperature in a Rigaku X-ray diffractometer.
The Rietveld analysis of the X-ray powder diffraction data (figure 1) confirmed that the sample is crystallized in the tetragonal phase with majority of the peaks indexed to the space group P4/nmm with lattice constants $a$ = $b$ = 4.001 {\AA} and $c$ = 8.35 {\AA}. There are some unindexed lines with small intensity which are arising from the impurity phases of Praseodymium oxide and CoAs. The procedure of preparation of NdCoPO was discussed in ref: 6.

\begin{figure}[h]
{\centering {\includegraphics[width=0.5\textwidth]{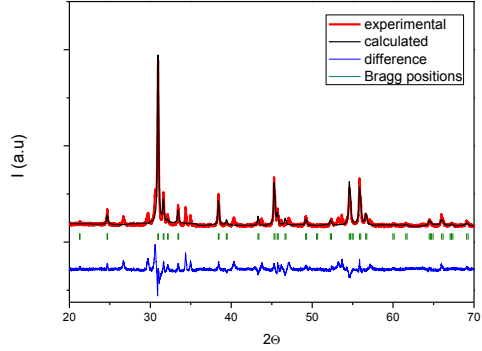}}\par} \caption{(Color online) Powder X-ray diffraction pattern of PrCoAsO at 300 K with Rietveld fit.} \label{structure}
\end{figure}

The $^{75}$As and $^{31}$P NMR measurements are performed on the powder samples of PrCoAsO and NdCoPO using a conventional phase-coherent spectrometer (Thamway PROT 4103MR) with a 7.0 T superconducting magnet (Bruker). The spectrum is recorded by changing the frequency step by step and recording the spin echo intensity by applying a $\pi/2- \tau - \pi/2$ solid echo pulse sequence. The temperature variation study is performed in an Oxford continuous flow cryostat equipped with a ITC503 controller. The spin lattice relaxation time ($T_1$) is measured using the saturation recovery method, applying a single $\pi$/2 pulse.

\section{$PrCoAsO$}

\subsection{$^{75}$As NMR spectra}

\begin{figure}[h]
{\centering {\includegraphics[width=0.5\textwidth]{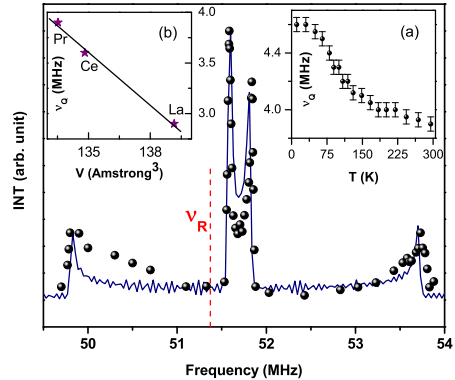}}\par} \caption{(Color online) $^{75}$As NMR spectra at 300 K in PrCoAsO. The continuous line corresponds to the theoretical fit. The vertical dashed line corresponds to the reference position of $^{75}$As nuclei at the field of 7T. Inset (a) shows temperature dependence of $\nu_Q$ (MHz) for PrCoAsO, inset (b) shows $\nu_Q$ (MHz) versus $V$ (\AA$^3$) for LaCoAsO \cite{Ohta}, CeCoAsO \cite{Rajib10} and PrCoAsO at 300 K.} \label{structure}
\end{figure}
\begin{figure}[h]
{\centering {\includegraphics[width=0.5\textwidth]{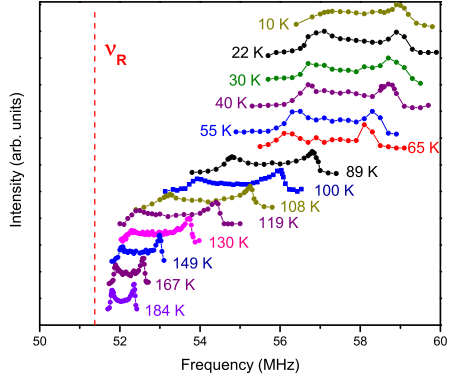}}\par} \caption{(Color online) $^{75}$As NMR spectra at different temperatures in PrCoAsO. The vertical dashed line corresponds to the reference position of $^{75}$As nuclei at the field of 7 T.} \label{structure}
\end{figure}

Figure 2 shows the $^{75}$As ($I$=3/2) NMR spectrum at 300 K in PrCoAsO which consists of a splitted central line and two satellites. The splitting of the central line is due to the combined effect of magnetic anisotropy and second order quadrupolar interaction. Assuming the principal axes of the electric field gradient (EFG) and magnetic shift tensors are parallel to each other, the spectrum can be simulated as shown by continuous line in figure 2.

Unlike the case of CeCoAsO \cite{Rajib10}, $^{75}$As NMR spectrum in PrCoAsO (figure 3) did not vanish below the FM transition temperature ($T_C$ $\sim$ 75 K). However, the magnetic anisotropy and the line width increase continuously. The increase in line width can be understood as due to the effect of the increase of magnetization. $^{31}$P NMR spectrum in LaCoPO \cite{Majumder09} was also detected below $T_C$. Whereas, in SmCoPO where $T_C$=110 K in a field of 7 T, the NMR line could not be detected below 130 K, due to the large increase in line width and anisotropy of the hyperfine field from below 300 K, originating mainly from Sm 4$f$ spin contribution \cite{Majumder12}. This indicates that in PrCoAsO the effect of 4$f$ moment on $^{75}$As NMR line shape is weaker than that in SmCoPO. The $^{75}$As NMR results in SmCoAsO are not yet reported. From figure 3 it is seen that no noticeable change in the $^{75}$As NMR line shape and position in PrCoAsO are discernable below 40 K, as expected from a change in the magnetic state, observed in $\mu$SR results below 50 K \cite{Sugiyama11}. A possible reason for this could be the suppression of the change in the local magnetic field below 50 K (observed from $\mu$SR study in zero magnetic field) in a field of 7 T, used in present NMR experiment. Recently from $\mu$SR, it was reported that the results of PrCoPO ($T_C$=48 K) are qualitatively identical to that of LaCoPO, despite the presence of magnetic moment of Pr$^{3+}$ ion \cite{Prando12}.

In the paramagnetic state of PrCoAsO, $^{75}$As nuclear quadrupolar coupling constant ($\nu_Q$) increases continuously below 300 K, (shown in inset (a) of figure 2) whose origin will be discussed later. This kind of increment of $\nu_Q$ was also observed in $^{75}$As NMR of CeCoAsO \cite{Rajib10}.
In LaCoAsO \cite{Michioka10}, CeCoAsO \cite{Rajib10} and PrCoAsO, the values of $\nu_Q$ at 300 K are 2.9, 3.6 and 3.9 MHz respectively indicating that the local EFG at the $^{75}$As site increases as we go from La to Pr. The distortion of the Co-As tetrahedra is related to decrement of the lattice volume from La to Pr as can be seen from the linear variation of $\nu_Q$'s (near room temperature) with the unit cell volume (inset (b) of figure 2).

\subsection{Knight shift and hyperfine field}

By fitting the experimental spectra with the simulated theoretical line we have estimated $K_c$, $K_{ab}$, $K_{iso}$, $K_{ax}$ and $\nu_Q$. $K_c$, $K_{ab}$ are the Knight shifts corresponding to $\theta$=0 and $\theta$=$\pi/2$, where $\theta$ is the angle between the direction of the external magnetic field and $z$ principle axis of the magnetic/electrostatic hyperfine interaction tensor. $K_{iso} = 2/3K_{ab} + 1/3K_{c}$ and $K_{ax} = 1/3(K_{c}-K_{ab})$. Figure 4 and 5 shows the temperature dependence of shift parameters $K_c$, $K_{ab}$, $K_{iso}$, $K_{ax}$ in PrCoAsO  down to 10 K and  for comparison, those reported in LaCoAsO \cite{Michioka10} and CeCoAsO \cite{Rajib10} have been included. The measured shift, $K = K_0 + K(T)$, where $K_0$ is the temperature independent contribution arising from conduction electron spin susceptibility, orbital susceptibility and diamagnetic susceptibility of core electrons. $K(T)$ arises from the temperature dependent susceptibility due to Co-3$d$ and Pr-4$f$ spins,
\begin {equation}
K(T) = (H_{hf}/N\mu_B)\chi(T)
\end {equation}
$H_{hf}$ is the total hyperfine field related to the total hyperfine coupling constant $A_{hf}$ by $H_{hf}$ = $A_{hf}$/$\gamma\hbar$g, $N$ is the Avogadro number and $\mu_B$ is the Bohr magneton. Inset of figure 5 shows the linear variation of $K_{iso}$ and K$_{ax}$ with $\chi = M/H$. The estimated values of $H_{hf}^{iso}$ and $H_{hf}^{ax}$ are 15.35 and -2.5 kOe/$\mu_B$ respectively.

Using the structural parameters for PrCoAsO we have calculated the dipolar field at the $^{31}$As nuclear site arising from Pr moment with the formula,
\begin{equation}
H_\textrm{dip} = \mu\sum\frac{(3r_jr_k - r^2\delta_{jk})}{r^5}; j,k=x, y, z
\end{equation}
$\mu$ is the magnetic moment of the Pr ion. The $H_{dip}^{ax}$ (-0.11 kOe/$\mu_B$) is one order of magnitude smaller than the experimental value of $H_{hf}^{ax}$. This indicates that the measured $H_{hf}^{ax}$ is governed mainly by the anisotropic hyperfine field produced by Pr-4$f$  and Co-3$d$ moments at the $^{75}$As nuclear site. So the observed values of $H_{hf}$ could be related to the effect of hybridization of the 4$s$ and  4$p$-orbitals of As with the 3$d$-orbitals of Co and 4$f$-orbitals of Pr via the conduction electrons.

\begin{figure}[h]
{\centering {\includegraphics[width=0.5\textwidth]{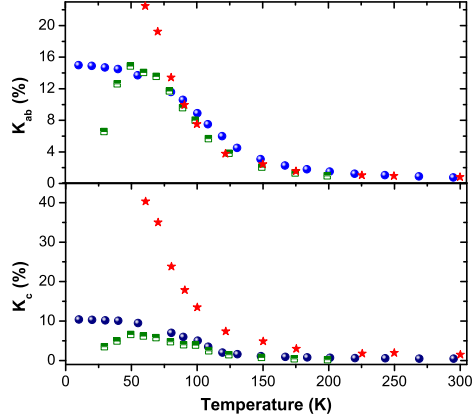}}\par} \caption{$K_c$ (\%), $K_{ab}$ (\%) versus temperature for LaCoAsO (star), CeCoAsO (half filled square) and PrCoAsO (filled circle). Data for LaCoAsO and CeCoAsO have been taken from \cite{Ohta} and \cite{Rajib10} respectively.} \label{structure}
\end{figure}
\begin{figure}[h]
{\centering {\includegraphics[width=0.5\textwidth]{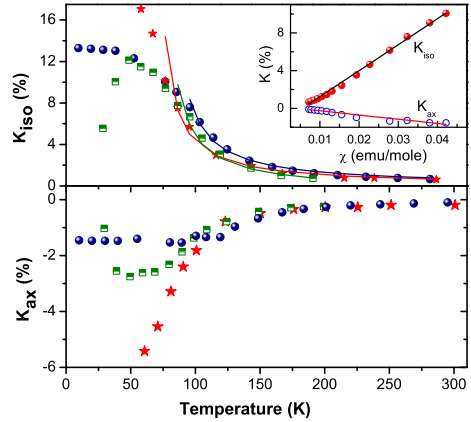}}\par} \caption{$K_{iso}$ (\%), $K_{ax}$ (\%) versus temperature for LaCoAsO (star), CeCoAsO (half filled square) and PrCoAsO (filled circle). Data for LaCoAsO and CeCoAsO have been taken from \cite{Ohta} and \cite{Rajib10} respectively. Inset shows $K_{iso}$, $K_{ax}$ versus $M/H$ for PrCoAsO.} \label{structure}
\end{figure}

Temperature dependence of $K_{iso}$ in the paramagnetic phase can be well described by the Curie-Weiss type behavior,
\begin {equation}
K_{iso}(T) = (H_{hf}^{iso}/N\mu_B)\frac{C}{(T-\theta)}
\end {equation}
as represented by the continuous lines in figure 5 for LaCoAsO, CeCoAsO and PrCoAsO. The estimated value of the effective moment,$P_{eff}$ from the Curie-Weiss constant ($C$) in LaCoAsO is 1.64$\mu_B$ with the Curie-Weiss temperature, $\theta$=70 K. The reduced value of Co effective moment with respect to the local Co moment (3.87$\mu_B$) indicates that Co 3$d$ electrons are of itinerant in nature. For CeCoAsO and PrCoAsO these parameters are 2$\mu_B$ (with $\theta$=75 K) and 2.18$\mu_B$ (with $\theta$=80 K) respectively. The $P_{eff}$ in case of PrCoAsO and CeCoAsO systems is due to the contribution of Co 3$d$ and rare earth 4$f$ i.e. $P_{eff} = \sqrt{(\mu_{eff}^d)^2 + (\mu_{eff}^f)^2}$. If we assume that for all other rare earths the $P_{eff}$ for Co 3$d$ is same then it is clear that in the paramagnetic state, there is also a contribution to $P_{eff}$ from 4$f$ electrons over that of the Co 3$d$ electrons.

$H_{hf}^{iso}$ and $H_{hf}^{ax}$ are found to vary linearly with the unit cell volume of LaCoAsO, CeCoAsO and PrCoAsO (figure 6). This indicates that the extent of hybridization of the electronic orbitals, involved in the hyperfine coupling constant, changes linearly with the change in the volume of the unit cell. Further $H_{hf}^{iso}$ also scales with $P_{eff}$ (figure 6). As the values of the spontaneous magnetization $M_s$ at $T$=0 and the $T_C$ are almost same in CeCoAsO and PrCoAsO \cite{Ohta09} the enhanced $P_{eff}$ values for CeCoAsO and PrCoAsO compared to that in LaCoAsO should be due to the rare earth-4$f$. Also the $P_{eff}$ values are less than those corresponding to the localized 4$f$ moment of Ce and Pr. This could arise due to the partial delocalization of 4$f$ electrons, due to a change in the band structure, resulting from the decrease in the unit cell volume for substitution of Ce and Pr in place of La.

\begin{figure}[h]
{\centering {\includegraphics[width=0.5\textwidth]{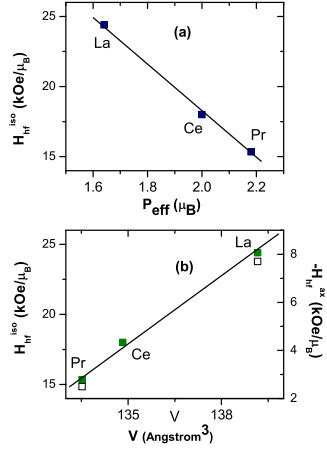}}\par} \caption{(Color online) (a) $H_{hf}^{iso}$ versus $V$ (\AA$^3$) and (b) $H_{hf}^{iso}$ (closed symbol) and $H_{hf}^{ax}$ (open symbol) versus $P_{eff}$ ($\mu_B$) for LaCoAsO \cite{Ohta}, CeCoAsO \cite{Rajib10} and PrCoAsO.} \label{structure}
\end{figure}

The values of $K_{ab}$ for LaCoAsO, CeCoAsO and PrCoAsO are almost same in the paramagnetic state but $K_c$ is lower in CeCoAsO and PrCoAsO compared to that in LaCoAsO which suggests that the local field produced at the $^{75}$As site along $c$-direction due to 4$f$ electrons, is of opposite sign compared to that of Co-3$d$. In case of CeCoAsO a peak in shift was observed near at 50 K, which was related to the crystal electric field (CEF) splitting of Ce 4$f$ and to the Schottky-type anomaly in specific heat \cite{Rajib10}. No such peak is observed in PrCoAsO. The present Knight shift data suggest that even at 80 K the $^{75}$As nucleus experiences a local magnetic field along $c$ axis due to the 4$f$ moments, because within the $ab$ plane, this field is same for La and Ce based analog, whereas, along c axis a drop in magnitude was observed. The inter layer AFM correlation of Co 3$d$ or Pr 4$f$ and intralayer FM correlation of Co 3$d$ moments can be the moment configuration. If this happens then in the Co-As layer the local field along c axis can be small due to the cancelation of AFM moments. This was also seen in NdCoAsO \cite{McGuire10} and SmCoPO \cite{Majumder12}. Further experiments like neutron elastic scattering have to be done to know the exact reason for this. The effect of CEF on the Pr-4$f$ levels can not be also neglected for the decrement of shift compared to the La analogue.

\subsection{Possible origin of the temperature dependence of quadrupolar coupling constant}

In case of PrCoAsO we have seen from $^{75}$As NMR that the $\nu_Q$ increases with the decrease of temperature (inset (a) of figure 2). The thermal (phonon) contribution can induce an increase in $\nu_Q$ at lower temperature. In this case, it follows the relation $\nu_Q^{phonon} = \nu_Q^0 - aT^{3/2}$ where $\nu_Q^0$ is the $\nu_Q$ at T = 0 K. It can easily be seen from the inset (a) of figure 2 that the temperature dependence of $\nu_Q$ does not follow this power law.

Most interestingly, we have seen that $\nu_Q$ scales with $K_{iso}$ or $\chi$ (figure 7). This type of scaling was seen in case of MnSi (itinerant ferromagnet) \cite{Yasuoka78}. To explain the linearity, a theory based on self consistent renormalization (SCR) approach was developed \cite{Takahashi78}.

In general, the total electric field gradient ($q$) in metals can be written as
\begin {equation}
q = \int_0^\infty dr (1+\gamma(r)) q(r)
\end {equation}
where $q(r)$ is the quadrupole charge density between $r$ and $r+dr$ and $\gamma (r)$ is its antishielding factor at $r$. The electric field gradient $q$ in metals consists of two contributions,
\begin {equation}
q = q_i + q_{el} (T)
\end {equation}
where $q_i$ is the ion core (lattice) electric field gradient and $q_{el}$ is the non-cubic part of the conduction electron charge distribution. One can easily evaluate the magnitude of $q_i$ by extrapolating the linear portion of $\nu_Q$ versus $\chi$ or $K_{iso}$ plot to $\chi$ or $K_{iso}$ = 0. A contribution to $q_{el}$ in terms of quadrupolar charge susceptibility, which can be influenced by spin-fluctuations through the mode-mode coupling between charge and spin density fluctuations, was calculated and for which a linear relation between $\nu_Q$ and $\chi$ was found \cite{Takahashi78}. Thus the linear relation between $\nu_Q$ and the intrinsic spin susceptibility $K_{iso}$ in PrCoAsO indicates a possible signature of the presence of the coupling between charge density and spin density fluctuations in PrCoAsO resulting a temperature dependent $\nu_Q$. Since $\nu_{Q;el}$($T$) is proportional to $\chi$ and not to magnetization, so it should not change with magnetic field. It is also necessary to do the NMR experiments in presence of different magnetic fields, in order to check that $\nu_Q$ is field independent, to discard the possibility of the presence of static magneto-elastic coupling which can change $q_i$ or $q_{el}$, as the magnetization increases with increasing magnetic field.

\begin{figure}[h]
{\centering {\includegraphics[width=0.5\textwidth]{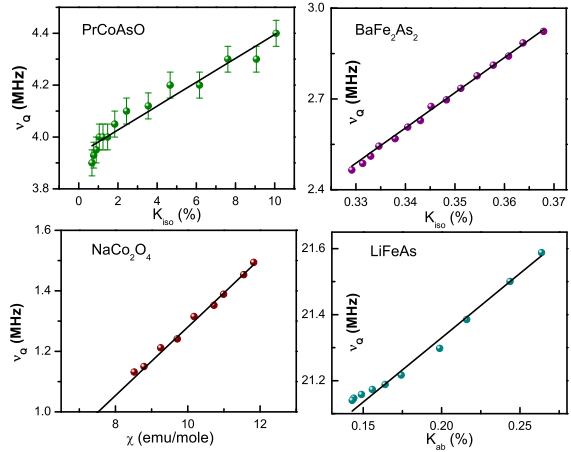}}\par} \caption{(Color online) $\nu_Q$ versus local susceptibility for PrCoAsO, BaFe$_2$As$_2$ \cite{Kitagawa08}, NaCo$_2$O$_4$ \cite{Ray99} and LiFeAs \cite{Baek11}.} \label{structure}
\end{figure}

We found that itinerant systems like BaFe$_2$As$_2$ \cite{Kitagawa08}, LiFeAs \cite{Baek11} and NaCo$_2$O$_4$ \cite{Ray99} also follow the same scaling relation (figure 7)  wherein the temperature dependence of $\nu_Q$ were also observed. In LiFeAs, the value of $^{75}$As $\nu_Q$ for $H$=0 T (from NQR) is 21.12 MHz (at 20 K) which is same when NMR has been done at 51.1 MHz \cite{Li10}. In BaFe$_2$AS$_2$, $^{75}$As NMR $\nu_Q$ = 3 MHz, at the resonance frequencies $\nu_R$ = 45 MHz \cite{Baek08} and 48.4MHz \cite{Kitagawa08}. This field independent value of $\nu_Q$ further supports the applicability of the above mentioned model to explain the $T$ dependent $\nu_Q$ in these systems.

\section{$NdCoPO$}

\subsection{$^{31}$P NMR spectrum, Knight shift and hyperfine field}
\begin{figure}[h]
{\centering {\includegraphics[width=0.4\textwidth]{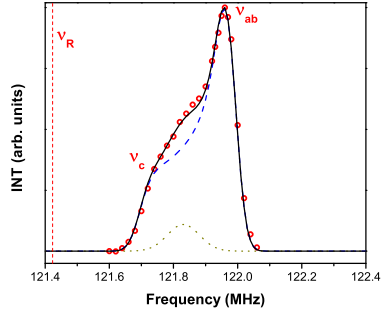}}\par} \caption{(Color online) $^{31}$P NMR spectrum of NdCoPO at 300 K. Continuous line is the total theoretical fit which consists of pure NdCoPO (dashed curve) and impurity (dotted curve). The vertical dashed line corresponds to the reference position of $^{31}$P nuclei at the field of 7 T.} \label{structure}
\end{figure}
Figure 8 shows the typical $^{31}$P NMR spectrum in polycrystalline NdCoPO at 300 K. The resonance line shape corresponds to a powder pattern of a spin I = 1/2 nucleus experiencing an axially symmetric local magnetic field, as expected for NdCoPO having tetragonal symmetry. The step in the low-frequency side corresponds to $H_0 \parallel c$ ($\theta = 0^\circ$) and the maximum in high frequency side corresponds to $H_0 \perp c$ ($\theta = 90^\circ$). The shift of the step with respect to the reference position ($\nu_R$), corresponds to $K_c$ and that of the maximum corresponds to $K_{ab}$. The continuous line superimposed on the experimental line is the calculated spectrum using Gaussian line shape along with the symmetric impurity line centered at 121.83 MHz. The separation between the step, $\nu_c$ and the maximum, $\nu_{ab}$ increases at low temperature along with line broadening.

Figure 9 shows the temperature dependence of $K_{iso}$ and $K_{ax}$ in NdCoPO in the temperature range 140 - 300 K (paramagnetic phase) along with those reported in LaCoPO \cite{Majumder09} SmCoPO \cite{Majumder12} for comparison. With the decrease of temperature, both the shift and the line width increase due to the increase of magnetization.

\begin{figure}[h]
{\centering {\includegraphics[width=0.5\textwidth]{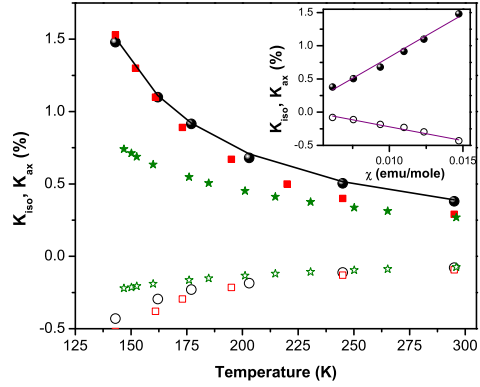}}\par} \caption{(Color online) Temperature dependence of $K_{iso}$ and $K_{ax}$. Filled circles, filled square, filled star corresponds to $K_{iso}$ for NdCoPO, LaCoPO \cite{Majumder09} and SmCoPO \cite{Majumder12} respectively. Open circles, open square, open star corresponds to $K_{iso}$ for NdCoPO, LaCoPO \cite{Majumder09} and SmCoPO \cite{Majumder12} respectively. Inset shows $K_{iso}$ and $K_{ax}$ versus $M/H$ for NdCoPO.} \label{structure}
\end{figure}

Inset of figure 9 shows the linear variation between $K_{iso}$ and $K_{ax}$ with $\chi$ which indicates unique isotropic and anisotropic hyperfine coupling constants. This further suggests that the presence of the impurity line in the NMR spectrum does not hamper the determination of the shift parameters of NdCoPO from theoretical fitting. Using equation 1 we obtain $H_{hf}^{iso}$ and $H_{hf}^{ax}$ to be 7.14 kOe/$\mu_B$ and -2.34 kOe/$\mu_B$ respectively which are of same order of magnitude as in LaCoPO and SmCoPO shown in figure 10. However, the $^{75}$As NMR hyperfine coupling constant derived in case of NdFeAsO and NdFeAsO$_{1-x}$F$_x$ were one order of magnitude higher than that in NdCoPO \cite{Jeglic09}. This means the hyperfine coupling becomes weaker when As is replaced by P and Fe is replaced by Co. This can happen if such replacement alters the band structure so that the overlap of the different orbitals involved in $H_{hf}$ is affected.

From the theoretical fitting of the $K_{iso}$ versus $T$ curve using equation 3, the estimated value of $P_{eff}$ obtained from the value of the Curie-Weiss constant ($C$) is 2.23$\mu_B$ with $\theta$=90 K for NdCoPO. For LaCoPO, $P_{eff}$ = 1.4$\mu_B$. This indicates that their is a contribution of Nd 4$f$ moment over that of Co-3$d$ moment on shift even in the paramagnetic state. In case of NdFeAsO$_{1-x}$F$_x$ the $P_{eff}$ value was close to local Nd-4$f$ moment where effect of Fe 3$d$ is negligible \cite{Jeglic09}. However, in case of NdCoPO if we assume that the Co-3$d$ moment contribution is same as in LaCoPO, then from the equation $P_{eff} = \sqrt{(\mu_{eff}^d)^2 + (\mu_{eff}^f)^2}$ one can easily see that the $P_{eff}$ value for Nd 4$f$ is lower than the local Nd 4$f$ moment, which reveals a signature of the itinerant character of Nd-4$f$ electrons in NdCoPO similar to that of the Co 3$d$ electrons.

Figure 10 shows that $H_{hf}^{iso}$ follows linear relationship with $P_{eff}$ from LaCoPO to SmCoPO. Whereas, $H_{hf}^{iso}$ and $H_{hf}^{ax}$ do not decrease linearly with the decrease of lattice volume similar to that in As based series (where the $^{75}$As NMR results are available only in LaCoAsO, CeCoAsO and the present results in PrCoAsO). In \emph{RE}CoPO series $H_{hf}^{iso}$ and $H_{hf}^{ax}$ decreases linearly with decrease of lattice volume from LaCoPO to NdCoPO. With further decrease of lattice volume in SmCoPO, both the coupling constants increase abruptly indicating a drastic change in the band structure. Therefore, the present results bring out the necessity of a systematic NMR study also in other members of \emph{RE}CoAsO and \emph{RE}CoPO series, in order to get a clear microscopic picture about the magnetic properties of both the series.

\begin{figure}[h]
{\centering {\includegraphics[width=0.5\textwidth]{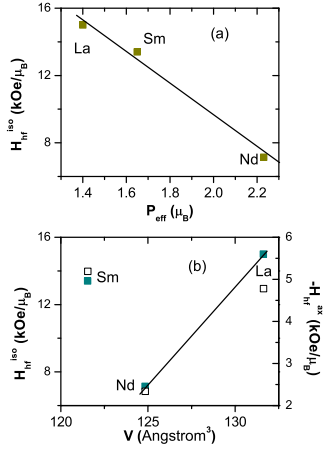}}\par} \caption{(Color online) (a) $H_{hf}^{iso}$ versus $P_{eff}$ ($\mu_B$) for LaCoPO \cite{Majumder09}, NdCoPO and SmCoPO \cite{Majumder12}, (b) $H_{hf}^{iso}$ (closed symbol) and $H_{hf}^{ax}$ (open symbol) versus $V$ (\AA$^3$).} \label{structure}
\end{figure}

\subsection{Nuclear spin-lattice relaxation rate ($1/T_1$)}
In general, ($1/T_1T)_{SF}$ is given by
\begin {equation}
(1/T_1T)_{SF} \propto \sum_q |H_{hf}(q)|^2\chi\prime\prime(q, \omega_n)/\omega_n
\end {equation}
where $\chi^{\prime\prime} (q, \omega_n$) is the imaginary part of the transverse dynamical electron spin susceptibility, $\gamma_n$ and $\omega_n$ are the nuclear gyromagnetic ratio and Larmor frequency respectively. $H_{hf}(q)$ is the hyperfine form factor.

Nuclear spin lattice relaxation time ($T_1$) is measured at the high intensity peak ($\nu_{ab}$ in figure 8) of $^{31}$P NMR spectrum at all the temperatures. Because of the presence of the impurity line near the position of the step ($\nu_{c}$) at almost all the temperatures, $T_1$ is not measured at this position, to avoid any discrepancy in the measured values. The temperature dependence of $(1/T_1T)_{ab}$ in NdCoPO is shown in figure 11 along with those reported in LaCoPO \cite{Majumder09}, and SmCoPO \cite{Majumder12} for comparison. It is interesting to note that the order of magnitude of relaxation time ($T_1$) in LaCoPO and NdCoPO are same. Whereas, in SmCoPO it is two orders of magnitude shorter. This indicates that in case of NdCoPO, the contribution to $T_1$ of the Nd-4$f$ electrons over that of Co 3$d$ electrons, is not as strong as it is in SmCoPO, even if the effective moment of Nd-4$f$ is higher than that of Sm-4$f$. The abrupt enhancement of $H_{hf}$ in SmCoPO (figure 10) should be the reason for the dominant role of RE-4$f$ electrons in $^{31}$P relaxation process of SmCoPO. So a comparison of the $^{31}$P $T_1$ data as well as the magnitude of the hyperfine coupling constants in LaCoPO, NdCoPO and SmCoPO clearly reveal a significant change in the electronic band structure in SmCoPO resulting an enhancement of $H_{hf}$ which shortens the spin-lattice relaxation time ($T_1$) along with the dominant effect of enhanced imaginary part of dynamic spin susceptibility.

In the paramagnetic state, the $P_{eff}$ value in NdCoPO is higher than that in SmCoPO still the value of $T_1$ in NdCoPO is much longer that in SmCoPO. This should be due to the decrement of the inter layer separation ($c_{SmCoPO} < c_{NdCoPO}$), possibly affecting the band structure, (which could enhance the $^{31}$P nuclear hyperfine coupling) and not due to the higher moment of Nd-4$f$ compared to that of  Sm-4$f$ electrons. From the inset of figure 11, it can be easily seen that $1/T_1T$ at 300K in NdCoPO is higher than that in LaCoPO due to the enhanced $P_{eff}$ value, but for SmCoPO it becomes higher though $P_{eff}$ value is small compared to NdCoPO. Moreover below the lattice volume of NdCoPO, the static magnetic properties (figure 10) as well as the dynamics of electrons changes drastically in case of SmCoPO. In the 1111 series Sm based systems always show interesting and different electromagnetic properties. In case of 1111 superconductors, SmFeAsO$_{1-x}$F$_x$ has highest $T_c$ and also the $^{19}$F relaxation rate is order of magnitude higher than that of La or Pr based analogues due to the strong role of Sm 4$f$ electrons \cite{Prando10, Majumder13}. The hybridization between Sm 4$f$ and Fe 3$d$ is stronger than that in the other compounds and as a result the Sm 4$f$ electrons are more itinerant in nature. The same case also happens in Co based magnetic compounds. Further the order of magnitude of $1/T_1T$ in SmCoPO \cite{Majumder12} and in SmFeAsO$_{1-x}$F$_x$ \cite{Majumder13, Prando10} is same when probed at $^{31}$P site and at $^{19}$F site respectively, which indicates that the role of Co 3$d$ or Fe 3$d$ are almost suppressed by the dominant effects of Sm 4$f$ electrons.

It would be interesting to probe the electromagnetic properties of the series Nd$_{1-x}$Y$_x$CoPO, where increment of Y concentration may result in the decrease of the lattice parameters. It can happen that for a particular concentration of Y, the lattice parameters become same as SmCoPO. This study can expected to reveal directly that only lattice parameter (not the moment of rare earth) is responsible  for the magnetic properties of these systems.

\begin{figure}[h]
{\centering {\includegraphics[width=0.5\textwidth]{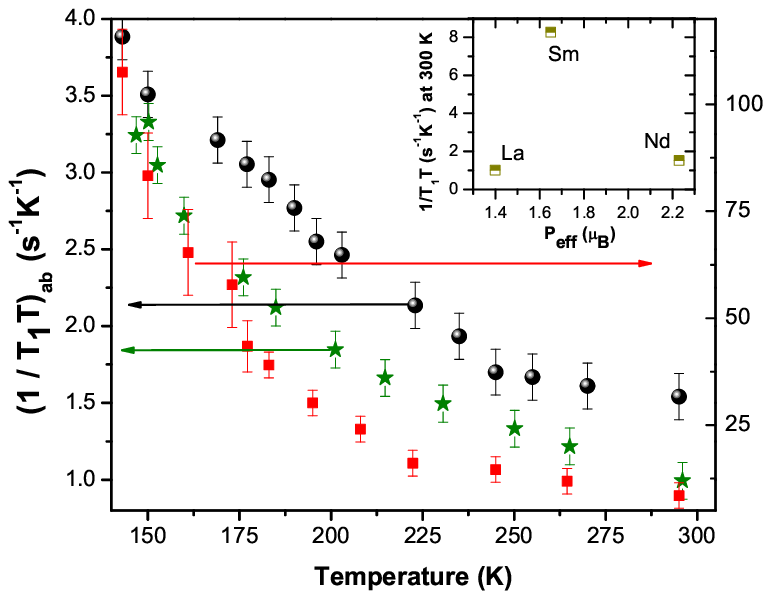}}\par} \caption{(Color online) Temperature dependences of $(1/T_1T)_{ab}$ for LaCoPO (star) \cite{Majumder09}, NdCoPO (circle) and SmCoPO (square) \cite{Majumder12}. Inset shows $1/T_1T$ versus $P_{eff}$ for LaCoPO \cite{Majumder09}, NdCoPO and SmCoPO \cite{Majumder12}.} \label{structure}
\end{figure}
\begin{figure}[h]
{\centering {\includegraphics[width=0.5\textwidth]{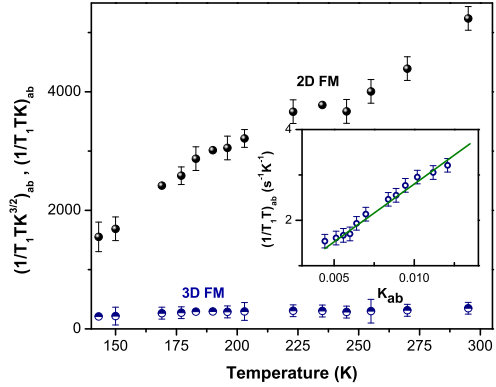}}\par} \caption{(Color online) $T$ versus $(1/T_1TK)_{ab}$ (half filled circle) and $(1/T_1TK^{3/2})_{ab}$ (filled circle) of NdCoPO. Inset shows $(1/T_1T)_{ab}$ versus $K_{ab}$ of NdCoPO.} \label{structure}
\end{figure}

When the Knight shift and the nuclear spin-lattice relaxation process are governed by conduction electrons, $1/T_1TK^2$ is constant. If there is an exchange interaction between the electrons then, using the Stoner approximation along with random phase approximation, modified Korringa relation can be written as \cite {Moriya63,Narath68,Lue99}
$S_0/T_1TK_{spin}^2 = \kappa(\alpha$), where $S_0 = (\hbar/4\pi
k_B)(\gamma_e/\gamma_n)^2$ and
\begin {equation}
\kappa(\alpha) = \langle(1-\alpha_0)^2/(1-\alpha_q)^2\rangle_{FS}.
\end {equation}
$\alpha_q = \alpha_0\chi_0(q)/\chi(0)$ is the $q$-dependent susceptibility enhancement, with $\alpha_0$ = 1 - $\chi_0(0)/\chi(0)$ representing the $q=0$ value. The symbol $\langle\rangle_{FS}$ means the average over $q$ space on the Fermi surface. $\chi$(0) and $\chi_0(q)$ represents the static susceptibility and the $q$ mode of the generalized susceptibility of non-interacting electrons respectively. $\kappa(\alpha)<$ 1 means the spin-fluctuations are enhanced around $q=0$, leading to the predominance of ferromagnetic correlations and $\kappa(\alpha) >$ 1 signifies that spin-fluctuations are enhanced away from $q=0$. This would indicate a tendency towards AF ordering (at $q\neq$0). In case of LaCoPO $\kappa(\alpha)<$ 1 both for $ab$ plane and $c$ direction, while for SmCoPO $\kappa(\alpha) >$ 1 for $ab$ plane and $\kappa(\alpha) <$ 1 for $c$ direction. In case of NdCoPO, $\kappa(\alpha)_{ab}$ $<$ 1 (0.04). This indicates a ferromagnetic correlations in the $ab$ plane.

\begin{figure}[h]
{\centering {\includegraphics[width=0.4\textwidth]{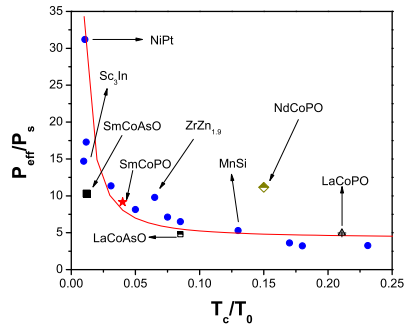}}\par} \caption{(Color online) $P_{\mathrm{eff}}/P_{\mathrm{s}}$ vs $T_\mathrm{C}$/$T_0$ plot (Rhodes-Wohlfarth plot). Solid line corresponds to Rhodes-Wohlfarth equation i.e. $P_{\mathrm{eff}}/P_{\mathrm{s}}$ $\propto$ ($T_\mathrm{C}$/$T_0$)$^{-3/2}$. The points (closed circle) are taken from \cite{Rhodes63}, LaCoPO from \cite{Majumder09}, SmCoAsO from \cite{Ohta}, SmCoPO from \cite{Majumder12}, LaCoAsO from \cite{Ohta09} and NdCoPO: present work.} \label{structure}
\end{figure}

According to the SCR theory of weak itinerant ferromagnet, if 3D/2D spin-fluctuations are dominant \cite {Hatatani95, Ishigaki98} then $1/T_1T$ $\propto$ $\chi^{1(3/2)}$. Figure 12 shows the $T$ versus $(1/T_1TK)_{ab}$ and $(1/T_1TK^{3/2})_{ab}$ plots revealing dominant 3D FM spin-fluctuations in the $ab$ plane of NdCoPO in the paramagnetic region in contrast to LaCoPO and SmCoPO where 2D FM correlations were present. It is to be pointed out that even if LaCoPO, NdCoPO and SmCoPO show a ferromagnetic transition for Co 3$d$ electrons, the dimensionality of electron spin fluctuations of Co 3$d$ are not same in all. In NdCoPO 3D spin fluctuations of Co 3$d$ dominates but it was 2D in case of LaCoPO and SmCoPO in the paramagnetic state.

If 3D FM spin fluctuation dominates the relaxation process then \cite{Corti07,Majumder10}
\begin {equation}
(1/T_1T) \simeq 3\hbar\gamma_n^2H_{hf}K/16\pi\mu_BT_0.
\end {equation}
where spin-fluctuation parameter $T_0$ characterizes the width of the spin excitations spectrum in frequency space. Hence by plotting (1/$T_1T)_{ab}$ versus $K_{ab}$ (inset of figure 12), we have calculated $T_0$ from its slope, which is 499 K, much smaller than that of LaCoPO and SmCoPO \cite{Majumder12}.

Further, according to the SCR theory of weak itinerant ferromagnet (WIF) \cite{Ohta09,Moriya73,Takahashi85,Moriya85}
\begin {equation}
T_C = (60c)^{-3/4}P_s^{3/2}T_A^{3/4}T_0^{1/4}
\end {equation}
where $c$=0.3353$\cdot\cdot\cdot$ and $P_{\mathrm{s}}$ is $M_{\mathrm{s0}}$ in $\mu_B$ unit. $T_A$ which characterizes the energy width of the dynamical spin-fluctuation spectrum and evaluated to be 19730 K (using equation 9) for NdCoPO.

This SCR theory is also used to evaluate the coefficient $\bar{F}_1$ (an important spin fluctuation parameter) of $M^4$ term, in the Landau expansion of free energy, which can be written as \cite{Takahashi86}
\begin {equation}
\bar{F}_1 = 4T_A^2/15T_0.
\end {equation}
This equation indicates that there is a significant renormalization effects due to the zero point spin-fluctuations, which modify expansion coefficients of the free energy with respect to the uniform magnetization and further confirms that at T = 0K, behavior of zero point spin-fluctuations have an important role in determining the magnetic properties in itinerant systems. $\bar{F}_1$ is 208028 K for NdCoPO (as estimated from equation 10) which is greater than that of SmCoPO (57562 K), This suggests a more localized character of the $d$ electrons in NdCoPO compared to that in SmCoPO.

According to the SCR theory, in case of weakly itinerant ferromagnets, zero point spin-fluctuation mainly contributes to the total spin-fluctuations and the thermal spin-fluctuations increase with temperature above $T_C$, which makes the value of the ratio of effective paramagnetic moment to the saturation moment, $P_{\mathrm{eff}}/P_{\mathrm{s}}$ greater than one. But in case of local ferromagnetic systems zero point spin-fluctuation is negligible and thermal spin-fluctuations are constant above $T_C$ and which gives $P_{\mathrm{eff}}/P_{\mathrm{s}}$ to be equal to one. Thus the ratio $P_{\mathrm{eff}}/P_{\mathrm{s}}$ $>$ 1 indicates that $P_{\mathrm{eff}}$ does not contributes to the static magnetic moment but to the dynamic magnetic response which indicates the role of spin-fluctuations in describing the magnetism of itinerant magnetic systems. From the above argument it is clear that the ratio $P_{\mathrm{eff}}/P_{\mathrm{s}}$ is important to distinguish between itinerant and local systems. Further due to the presence of the spin-fluctuation in itinerant systems, the spin-fluctuation parameter $T_0$ is also of interest and according to the SCR theory, $T_C/T_0$ = 1 in case of localized ferromagnet, but as the itinerant character sets in, the value becomes less than one. It has been seen that all itinerant ferromagnetic materials follow a universal plot (Rhodes-Wohlfarth plot) between $P_{\mathrm{eff}}/P_{\mathrm{s}}$ and $T_C/T_0$ \cite{Takahashi86,Rhodes63}. To check the validity of the obtained spin fluctuation and thermodynamic parameters in NdCoPO and SmCoPO, we have plotted $P_{\mathrm{eff}}/P_{\mathrm{s}}$ versus $T_C/T_0$ in figure 13. Though NdCoPO and SmCoPO are not purely itinerant ferromagnet and also there are contributions of 4$f$ electrons on Knight shift and relaxation rate $(1/T_1)$ over Co 3$d$, we have tried to see whether NdCoPO and SmCoPO follow RW plot or not. From figure 13 one can see that NdCoPO and SmCoPO roughly follow RW relation $P_{\mathrm{eff}}/P_{\mathrm{s}}\sim(T_\mathrm{C}/T_0)^{-3/2}$. From the plot one can say that the delocalization character of Co 3$d$ electrons increases from LaCoPO to NdCoPO to SmCoPO. For comparison, we have also included SmCoAsO \cite{Ohta} in the RW plot, which is close to SmCoPO, indicating the change of the itinerant character of Co 3$d$ is nominal for the replacement of As by P.

\section{CONCLUSION}

We have reported $^{75}$As and $^{31}$P NMR in PrCoAsO and  NdCoPO respectively. The similarities and dissimilarities of the NMR results in the As and P based systems with different rare earths, are discussed. The role of 3$d$ electrons, in the Co and Fe based systems, has also been revealed indicating an interesting interplay between 4$f$ and 3$d$ in these magnetic systems.

We have proposed a possible spin structure in PrCoAsO from $^{75}$As NMR results. A scaling between $\nu_Q$ and K$_{iso}$ or $\chi$ indicates the presence of a coupling between charge density and spin density fluctuations as proposed theoretically by Moriya for itinerant magnetic systems. It is shown that similar scaling also holds in the itinerant systems e.g. BaFe$_2$As$_2$ \cite{Kitagawa08}, NaCo$_2$O$_4$ \cite{Ray99} and LiFeAs \cite{Jeglic09}.

The behavior of Knight shift and spin lattice relaxation data in the paramagnetic state of NdCoPO was compared with those reported in LaCoPO and SmCoPO. The role of RE-4$f$ effective moment is less prominent in NdCoPO. These results suggest a  comparatively large modification of the band structure in SmCoPO, resulting from the decrement of inter layer separation. Nuclear spin lattice relaxation rate $(1/T_1$) in NdCoPO at 300 K, is nearly isotropic. The relaxation data was analyzed using the SCR theory of itinerant ferromagnet and was shown that, along the $ab$ plane, three dimensional ferromagnetic correlations are present among the Co-3$d$ spins. Moreover, the different spin fluctuation parameters were evaluated. We conclude that the role of Co 3$d$ electrons is not universal for different members of the family in the paramagnetic state along the $ab$ plane.

\end{document}